\documentclass[twocolumn,aps,prb,superscriptaddress,showpacs,floatfix]{revtex4}
\usepackage{epsfig,graphicx,times}
\newcommand{\mb}[1]{\mathbf{#1}}
\newcommand{\ket}[1]{\left| #1\right\rangle}   
\newcommand{\bra}[1]{\left\langle #1 \right|} 

\begin{document}

\title{An Efficient Algorithm for Density Functional Theory Simulation of Large Quantum
Dot Systems}

\author{Hong Jiang}
\affiliation{Department of Chemistry, Duke University, Durham, North Carolina 27708-0354}
\affiliation{Department of Physics, Duke University, Durham, North Carolina 27708-0305}
\affiliation{College of Chemistry and Molecular Engineering, Peking University, Beijing, China 100871}
\author{Harold U. Baranger}
\email[]{baranger@phy.duke.edu}
\affiliation{Department of Physics, Duke University, Durham, North Carolina 27708-0305}
\author{Weitao Yang}
\email[]{weitao.yang@duke.edu}
\affiliation{Department of Chemistry, Duke University, Durham, North Carolina 27708-0354}

\date{\today{}}

\begin{abstract}
Kohn-Sham spin-density functional theory provides an efficient and accurate model
to study electron-electron interaction effects in quantum dots,
but its application to large systems is a challenge. An efficient
algorithm for the density-functional theory simulation of quantum dots is
developed, which includes the particle-in-the-box representation of the Kohn-Sham orbitals,
an efficient conjugate gradient method to directly minimize the total energy,
a Fourier convolution approach for the calculation of the Hartree potential,
and a simplified multi-grid technique to accelerate the convergence. The new algorithm
is tested in a 2D model system. Using this new algorithm, numerical studies of large
quantum dots with several hundred electrons become computationally affordable.
\end{abstract}
\maketitle

\section{Introduction}
Quantum dots (QD) are one kind of nano-device in which the motion of electrons
is quantized in all three dimensions through the lateral confinement of a
high-mobility modulation-doped two-dimensional electron gas in a semiconductor
hetero-structure.\cite{Kouwenhoven97,Alhassid00RMP,Aleiner02} Both quantum
interference and electron-electron interactions play important roles, and their
interplay underlies many properties that are fascinating in terms of both
fundamental mesoscopic physics and future technical applications. Various
theoretical models have been developed to explain experimental discoveries and
to predict new properties. The Kohn-Sham (KS) density functional theory (DFT)
method \cite{ParrYang89,Jones89RMP,DreizlerGross90} provides an accurate
numerical model to study electron-electron interaction effects in QD systems.
\cite{Kumar90, Macucci93,Jovanovic94, Stopa96, Nagaraja97,Koskinen97,Lee98,
Bednarek01,Yakimenko01,Pi01,Hirose02, Jiang02} In spite of its comparatively
low computational cost, previous DFT calculations are limited to systems in
which the electron number is generally less than a few tens. In many
experimental cases, however, the electron numbers involved are more than
several hundred. The main theoretical approaches in the regime of large number
of electrons are statistical methods \cite{Alhassid00RMP, Aleiner02, Ullmo01b,
Usaj02} which are usually based on some general assumptions whose validity, in
many cases, is yet to be justified. It is therefore desirable to explore the
large dot regime directly by using numerically more accurate models such as
DFT. This imposes a demanding computational task because to obtain meaningful
statistics, many calculations with several hundred electrons need to be done.
It is therefore compelling to develop more efficient numerical techniques for
DFT simulation of QD systems; this is the main object of the current study.

Two key issues are involved in the numerical implementation of the KS-DFT
method:\cite{Payne92,Kresse96} (1) the numerical representation
 of wave functions and the KS Hamiltonian, and
(2) the solution of the numerical KS equation. While local basis sets (mainly
Gaussian-type orbitals) dominate in conventional quantum chemistry of molecular
systems, the plane wave (PW) basis,\cite{Payne92,Kresse96} combined with the
pseudo-potential method, is widely used in the \textsl{ab initio} electronic
calculations of various material systems, in which the fast-Fourier transform
(FFT) method can be used to take full advantage of the periodicity of crystal
structures. In principle, the PW method is valid only for periodic systems, but
aperiodic systems can be treated by introducing the super-cell
technique.\cite{Payne92} In recent years, several groups have been advocating
the use of basis-free real space methods for electronic structure calculations
of finite systems, in which the wave functions are represented in real space,
and the kinetic energy operator is discretized by a high-order finite
difference (FD) method. \cite{Chelikowsky94} With a given representation, there
are various methods to solve the resultant numerical KS equation. Roughly they
fall into two different types: methods that minimize the total energy directly,
and those that solve the KS equation in a self-consistent way. In addition, for
DFT simulations of finite systems, another important issue is the calculation
of the Hartree potential.

Aiming at modelling QD systems efficiently, we have developed new techniques in
both the numerical representation of KS orbitals and the solution of the KS
equation. We note that in QD systems, the wave functions vanish at the
boundaries, and in some cases even the hard-wall boundary condition is used.
For a function in a rectangular box with zero boundary values, the most natural
basis set is the particle-in-the-box (PiB) basis set. The kinetic energy
operator is diagonal in the PiB space and the transformation between real and
PiB space can be efficiently performed by the fast sine-transform (FST) method,
which is a variant of the FFT. For the solution of the KS equation, we modified
Teter, Payne and Allan (TPA)'s band-by-band conjugate gradient method
\cite{Teter89} to get a more efficient direct minimization approach.

The outline of the paper is as follows. In the next section, after
a simple description of the KS-DFT method, we present the main
components of our algorithm: (1) the PiB representation of the KS
equation; (2) a modified band-by-band conjugate gradient method
for the direct minimization of the KS total energy; (3) a Fourier
convolution method for the calculation of the Hartree potential
that was developed by Martyna and Tuckerman \cite{Martyna99} in
the PW pseudo-potential calculations; and (4) a simplified
multi-grid technique to accelerate the convergence. In Section
III,  the new algorithm is tested in a 2D model QD system with
electron number $N=100$. Section IV summarizes the main results
and concludes the paper.

\section{Method}
\subsection{Kohn-Sham spin-density-functional theory}
Considering the important role played by electron spin in QD systems, we take
the effect of spin polarization explicitly into account in the framework of
Kohn-Sham spin-density functional theory (KS-SDFT).
\cite{ParrYang89,DreizlerGross90} In KS-SDFT, the ground state energy of an
interacting system with electron number $N$ and the total spin $S$ in the local
external potential \( V_{\mathrm{ext}}(\mathbf{r}) \) is written as a
functional of spin densities \( n^{\sigma } \) with \( \sigma =\alpha, \beta \)
denoting spin-up and spin-down, respectively,
\begin{eqnarray}
E\left[ n^{\alpha },n^{\beta }\right] =T_{s}\left[ n^{\alpha },n^{\beta }\right]
+\int n(\mb{r})V_\mathrm{ext}(\mb{r})d\mb{r} &  & \nonumber \\
+\frac{1}{2}\int \frac{n(\mb{r})n(\mb{r}')}{\left| \mb{r}-\mb{r}'\right|
}d\mb{r}d\mb{r}' +E_{\mathrm{xc}}\left[ n^{\alpha },n^{\beta }\right]  &.
\label{eq: E0}
\end{eqnarray}
(Effective atomic units are used through the paper: for GaAs-AlGaAs QDs, the
values are 10.08 meV for energy and 10.95 nm for length.)
 \( T_{s}\left[ n^{\alpha },n^{\beta }\right] \) is the kinetic energy of
the KS non-interacting reference system which has the same ground state spin
density as the interacting one, and \( E_\mathrm{xc}\left[ n^{\alpha },n^{\beta
}\right]  \) is the exchange-correlation energy functional  The spin densities
$n^{\sigma }$ satisfy the constraint $\int
n^{\sigma}(\mb{r})d\mb{r}=N^{\sigma}$ with $ N^\alpha=(N+2S)/2$ and $N^\beta =
(N-2S)/2$.

Assuming that the ground state of the non-interacting reference system is
non-degenerate, the non-interacting kinetic energy is given by $T_s\left[
n^{\alpha },n^{\beta }\right]=\sum _{i,\sigma } \left\langle \psi _{i}^{\sigma
}\left| -\frac{1}{2}\nabla ^{2}\right| \psi _{i}^{\sigma }\right\rangle $, and
the ground state spin density is uniquely expressed as
\begin{equation}
n^{\sigma }(\mb{r})=\sum _{i}^{N^\sigma}\left| \psi _{i}^{\sigma }(\mb{r})\right| ^{2},
\quad \sigma=\alpha,\beta.
\end{equation}
Here $\psi_{i}^{\sigma}$ are the lowest single-particle orbitals which are obtained
from
\begin{equation}
\label{eq:KS}
\mb{H}^{\sigma }_\mathrm{KS}\psi _{i}^{\sigma }(\mb{r})=\varepsilon _{i}^{\sigma }
\psi _{i}^{\sigma }(\mb{r}),
\end{equation}
with the KS Hamiltonian $\mb{H}_{\mathrm{KS}}$ defined as
\begin{equation}
\label{eq:HKS}
\mb{H}_{\mathrm{KS}}^{\sigma }\equiv -\frac{1}{2}\nabla ^{2}+V_\mathrm{ext}(\mb{r})+V_{H}[n;\mb{r}]
+V_{\mathrm{xc}}^{\sigma }[n^{\alpha },n^{\beta };\mb{r}].
\end{equation}
\( V_{H}[n;\mb{r}] \) and \( V_{\mathrm{xc}}^\sigma[n^{\alpha },n^{\beta
};\mb{r}]\) are the Hartree and exchange-correlation potentials, respectively,
\begin{eqnarray}
\label{eq:VH}
V_{H}[n;\mb{r}]& \equiv &\int \frac{n(\mb{r}')}{|\mb{r}-\mb{r}'|}d^{3}\mb{r}', \\
V_{\mathrm{xc}}^\sigma[n^{\alpha },n^{\beta }; \mb{r}] & \equiv & \frac{\delta E_\mathrm{xc}
\left[ n^{\alpha} ,n^{\beta }\right]}{ \delta n^{\sigma}(\mb{r})}.
\end{eqnarray}

We have used the local spin-density approximation
(LSDA)\cite{ParrYang89,DreizlerGross90} for $E_\mathrm{xc}$, which is widely
used for the modelling of material systems. Although more accurate
exchange-correlation functional forms such as the generalized-gradient
approximation (GGA) are available, it has been shown that the GGA results are
close to those from LSDA calculations in QD systems. \cite{Lee98} In terms of
the implementation, the calculation of $V_\mathrm{xc}$ is trivial when the spin
densities are in real space as is the case in our algorithm, but the
calculation of \( V_{H} \) requires more efforts as will be shown later.

\subsection{PiB Representation}
To simplify the notation, we will take 1-D systems as an example,
the generalization to higher dimensional cases being straightforward.
Any regular function \( f(x) \) that is localized in the finite region
\( 0<x<L \) with zero boundary values can be expanded as
\begin{equation}
 f(x)=\sum_n C_n \sqrt{\frac{2}{L}} \sin\frac{n\pi x}{L}
\end{equation}
and the expansion coefficients $C_{n}$ are
\begin{equation}
C_n =\sqrt{\frac{2}{L}}\int _{0}^{L}f(x)\sin
\frac{n\pi x}{L}dx.\label{eq:coef}
\end{equation}
 Integrating Eq. (\ref{eq:coef}) numerically on a set of equally
spaced discrete points \{\( x_{j}\equiv j\Delta x \equiv j\frac{L}{N_x}\}\) using the
extended trapezoidal formula \cite{NumRecip} leads to
\begin{equation}
\label{eq:coef2}
C_{n}=\sqrt{\frac{2}{L}}\Delta x\sum _{j=1}^{N_x-1}f_{j}\sin \frac{\pi jn}{N_x}
=F_{n}\frac{\sqrt{2L}}{N_x}
\end{equation}
with \( f_{j}\equiv f(x_{j}) \) and
\begin{equation}
\ F_{n}  \equiv  \sum _{j=1}^{N_x-1}f_{j}\sin \frac{\pi jn}{N_x}  \equiv \mb{FST}\{f_j\}
\end{equation}
where $\mb{FST}\{f_j\}$ denotes the fast sine-transform of the data $\{f_j\}$.

One of the key ingredients in our algorithm is the action of the single-particle
Hamiltonian operator on wave functions,
\begin{equation}
\label{eq:Hf}
\mb{H}f(x)=\mb{T}f(x)+\mb{V}f(x),
\end{equation}
the efficiency of which is critical for the performance of the whole
algorithm. Whereas the potential energy operator is diagonal in real space,
the kinetic energy operator is diagonal in PiB space. Wave functions
can be transformed between the two spaces efficiently by the fast sine
transform. In our algorithm, wave functions are in discrete real space
\( \{f_{j}\} \), and the application of the potential energy operator
is therefore trivial, \( \mb{V}\{f_{j}\}=\{V_{j}f_{j}\} \), where $V_j$ is
the value of the potential at the point $x_j$. The kinetic
energy operator is applied to wave functions in PiB space,
\begin{equation}
\label{eq:Tf1}
\mb{T}f(x)\equiv -\frac{\hbar ^{2}}{2m}\nabla ^{2}f(x)=\frac{2}{N_x}
\sum _{n}F_{n}\frac{\hbar ^{2}k_{n}^{2}}{2m}\sin {\frac{n\pi x}{L}},
\end{equation}
with \( k_{n}=n\pi/L \), which in the discrete form becomes
\begin{equation}
\label{eq:Tf2}
\mb{T}\{f_{j}\}=\frac{2}{N_x}\mb{FST}\{F_{n}\frac{\hbar ^{2}k_{n}^{2}}{2m}\}.
\end{equation}

The PiB representation formulated above is closely related to the PW method.
In fact, for finite systems with soft-wall boundaries, the two representations
are numerically equivalent. Mathematically, however, they are different:
The PiB basis set is real and the zero boundary condition is imposed by
the basis set itself, but in the PW case the basis functions are complex
and the wave functions are forced to be
zero at the boundaries by the external potential of systems under study.
The PW method will fail in the case of the hard-wall boundary problem, which is quite
common in the studies of quantum dots, but the PiB method is still valid.

\subsection{Direct-minimization conjugate gradient method for the Kohn-Sham equation}
In direct minimization approaches, the KS total energy functional
is minimized directly over orbital wave functions under
orthonormal constraints. We have made several important
modifications to TPA's band-by-band conjugate-gradient scheme to
obtain higher efficiency. Here we give an outline of the
algorithm, and emphasize the modifications we have made.

The basic idea of a band-by-band scheme is to minimize the total energy over
one band (or orbital) at a time, which, compared to other conjugate-gradient
schemes,\cite{Stich89,Gillan89} has the following advantages: (1) much lower
requirement for storage space, (2) simpler implementation, and (3) particularly
in our algorithm, an efficient approximate line-minimization scheme, as will be
shown.

First the steepest descent (SD) vector for the $i$-th orbital at the $m$-th
iteration is calculated from
\begin{equation}
\ket{\zeta ^{m}_{i}} =\left( 1-\sum _{j\neq i}\ket{ \psi_{j}}\bra{\psi_{j}} \right)
(\lambda _{i}^{m}-\mb{H}_\mathrm{KS})\ket{\psi_{i}^{m}}
\end{equation}
with \( \lambda _{i}^{m}=\bra{\psi_{i}^{m}}\mb{H}_\mathrm{KS}\ket{\psi _{i}^{m}} \)
 (to simplify the notation, we use Dirac's state vector notation and
drop the spin index in the following formulation).
In TPA's algorithm, the SD vector
is preconditioned before it is used to build the conjugate vector. In our calculations, however,
it was found that although in many cases preconditioning
does accelerate the convergence, its effect is not always positive.
On the other hand, the computational overhead in the preconditioning step, which
involves another orthogonalization process as well as the action of the preconditioning
operator on the SD vector, can be expensive for large systems. As shown later,
by using a simplified multi-grid technique, we can achieve fast convergence
even without preconditioning of the SD vectors.

The conjugate vector \( \ket{\varphi _{i}^{m} } \) is then constructed
as a linear combination of the SD vector \( \ket{\zeta_{i}^{m}} \)
and the previous conjugate vector,
\begin{equation}
\ket{\varphi_i^{m}} =\ket{\zeta_i^{m}} +\gamma_i^{m} \ket{\varphi_i^{m-1}}
\end{equation}
where
\begin{equation}
\gamma _{i}^{m}=\frac{\left\langle \zeta _{i}^{m}\right. \left|
\zeta _{i}^{m}\right\rangle }{\left\langle \zeta _{i}^{m-1}\right.
\left| \zeta _{i}^{m-1}\right\rangle }
\end{equation}
with \( \gamma _{i}^{1}=0 \). The conjugate vector is further orthogonalized
to the present band $\ket{\psi_i^m}$ and normalized ($\mathbf{N}$ is denoted
as the normalization operator),
\begin{equation}
\ket{\varphi_i^{\prime m}} =\mathbf{N}\left( 1-\ket{\psi_{i}^{m}}
\bra{\psi_{i}^{m}}\right) \ket{\varphi_i^m}
\end{equation}
The new wave function for the $i$-th orbital \( \ket{\psi ^{m+1}_{i}}  \) is
formed from the linear combination
\begin{equation}
\ket{\psi_i^{m+1}} =\ket{\psi_i^{m}} \cos \theta_\mathrm{min}
+\ket{\varphi_i^{\prime m}}\sin \theta _\mathrm{min}\label{eq:update}
\end{equation}
which is guaranteed to remain normalized and orthogonal to all other orbitals.
\( \theta _\mathrm{min} \) is obtained by minimizing the total energy as a
function of \( \theta  \) with \( \ket{\psi_{i}(\theta)}
=\ket{\psi_{i}^{m}}\cos\theta+\ket{\varphi_i^{\prime m}}\sin\theta  \). In
TPA's algorithm, \cite{Teter89,Payne92} \( \theta _\mathrm{min} \) is
determined by the following approximate scheme: The total energy as a function
of $\theta$ is approximated by
 \( E(\theta )\approx E_{avg}+A_{1}\cos {2\theta }+B_{1}\sin {2\theta } \); the three
unknowns, $E_{avg}$, $A_1$ and $B_1$, are determined according to three pieces of
information: $E(\theta=0)$,
$\frac{\partial E(\theta)}{\partial \theta}|_{\theta=0}$ and $E(\theta=\pi/300)$.

Here we propose a more efficient approximate scheme for the determination of \( \theta _\mathrm{min} \).
The derivative of \( E(\theta ) \) with respect to $\theta$ can be obtained from
\begin{eqnarray}
\lefteqn{ \frac{\partial E(\theta )}{\partial \theta }
 =  2\bra{\varphi_i^{\prime m}} \mb{H}_\mathrm{KS}(\theta )\ket{\psi _{i}^{m}}\cos 2\theta }\nonumber \\
& & - \left( \bra{\psi_{i}^{m}} \mb{H}_\mathrm{KS}(\theta )\ket{\psi _{i}^{m}} -\bra{\varphi_i^{\prime m}}
\mb{H}_\mathrm{KS}(\theta )\ket{\varphi_i^{\prime m}} \right) \sin 2\theta \nonumber \\
\end{eqnarray}
Assuming \( \mb{H}_\mathrm{KS}(\theta )\approx \mb{H}_\mathrm{KS}(0) \),
from \( \frac{\partial E(\theta )}{\partial \theta }=0 \) we get
\begin{equation}
\theta _\mathrm{min}=\frac{1}{2}\tan ^{-1}\frac{B}{A}
\end{equation}
 with
\begin{eqnarray}
A=-\bra{\psi_{i}^{m}} \mb{H}_\mathrm{KS}(0)\ket{\psi _{i}^{m}}
+\bra{\varphi_{i}^{\prime m}}\mb{H}_\mathrm{KS}(0)\ket{\varphi_{i}^{\prime m}}
\end{eqnarray}
 and
\begin{equation}
B=2\bra{\varphi_i^{\prime m}} \mb{H}_\mathrm{KS}(0)\ket{\psi _{i}^{m}}.
\end{equation}
The underlying approximation in our scheme is similar to that of
TPA's, but our scheme is much more efficient in terms of
computational effort: In TPA's scheme, at each band iteration the
total energy must be calculated twice, i.e. $E(\theta=0)$ and
$E(\theta=\pi/300)$; in our scheme, the most time-consuming step
is the action of $\mb{H}_\mathrm{KS}$ on $\ket{\varphi_i^{\prime
m}} $, which is much faster than the calculation of the total
energy.  Considering further the fact that the total number of
band iterations can be very large, we note that an efficient
line-minimization scheme such as ours is crucial to reduce the
computational effort.

In TPA's algorithm, after the wave-function is updated according to
Eq. (\ref{eq:update}), the Kohn-Sham Hamiltonian is updated immediately,
which involves the reconstruction of \( V_{H}(\mb{r}) \) and \( V_{\mathrm{xc}}(\mb{r}) \)
according to the new density. This is actually quite expensive for large systems.
On the other hand, we expect that the KS Hamiltonian will not experience
large changes inside the iterations of a single orbital. So in our algorithm, we update
\( \mb{H}_\mathrm{KS} \) after every $N_\mathrm{update}$ band iterations, and the optimal value of
\( N_\mathrm{update} \) will be explored in the next section.

In each orbital, the procedure described above is repeated $N_\mathrm{band}$ times;
the iterations are then started on the next orbital. After the wave functions
of all orbitals are updated in this way, the total energy is calculated and is compared
to that of the previous cycle to determine if the final convergence is achieved.
The main parameters in the algorithm are \( N_\mathrm{band} \) and \( N_\mathrm{update} \).
Their effects on the performance of the algorithm will be tested in detail
in the next section.

\subsection{Calculation of \protect\( V_{H}\protect \)}
For finite systems, the simplest and perhaps most inefficient way to calculate
\( V_{H} \) is by direct numerical integration, which is feasible only for
small systems. Another widely-used approach is to solve the Poisson equation
equivalent to Eq. (\ref{eq:VH}). Though the Poisson equation itself can be
solved with great efficiency, the calculation of boundary values can be quite
expensive even by using efficient multipole expansion techniques. Additionally,
the Poisson solver approach is valid only for 3D systems; in the case of 2D
systems, there is no Poisson equation equivalent to Eq. (\ref{eq:VH}). In
recent years, several schemes have been proposed to extend the conventional
Fourier convolution method to finite
systems.\cite{Onida95,Fraser96,Jarvis97,Martyna99} In particular we have
incorporated Martyna and Tuckerman's method\cite{Martyna99} into our algorithm.
Considering that the Martyna-Tuckerman method was developed mainly for the
modelling of molecular and material systems within the plane-wave
pseudo-potential framework, we will formulate the approach here with some
detail.

The calculation of the Hartree potential is straightforward for periodic
systems, but this is not the case for finite aperiodic systems.  The potential
$V_H(\mb{r})$ has the form of the convolution between the density and the
Coulomb interaction kernel, $v_c(\mb{r})=1/r $, which has the following simple
relation in the Fourier space,\cite{NumRecip}
\begin{equation}
\tilde{V}_{H}(\mb{k})=\tilde{n}(\mb{k}) \tilde{v}_c(\mb{k}),
\label{eq:VH2}
\end{equation}
where $\tilde{f}(\mb{k})$ refers to the Fourier transform of $f(\mb{r})$. Eq. (\ref{eq:VH2})
is useful only when we have the analytical form of $\tilde{n}(\mb{k})$ and can perform the
inverse Fourier transform of $\tilde{V}_H(\mb{k})$ analytically, which is not true for most cases
where the density is usually represented in discrete real space.

\begin{figure}
\includegraphics[width=2.8in]{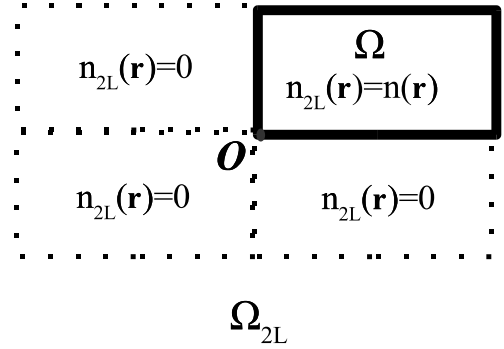}
\caption{\label{fig:VH} Illustration of the Fourier convolution
method for the calculation of the Hartree potential in finite
systems. The original grid where the density and potential are
defined is extended into doubled space. The density in the
extended region is set to zero. Imposing the periodic boundary
condition for this extended grid, the unphysical interaction
between neighboring super-cells can be avoided.}
\end{figure}

When applying the Fourier method to discrete finite systems,
periodic boundary conditions are always assumed. It has long been
known that the unphysical interactions between neighboring
super-cells can be avoided by calculating \( V_{H} \) in a doubly
extended grid. \cite{Hockney70} In particular for 2D systems as
illustrated in Fig. \ref{fig:VH}, the original \( L_{x}\times
L_{y} \) grid ($\Omega$) is extended to \( 2L_{x}\times 2L_{y} \)
($\Omega_{2L}$). The density in the extended grid is defined as
\begin{equation}
n_{2L}(\mb{r})=\left\{ \begin{array}{ll}
n(\mb{r}) &\textrm{if $\mb{r} \in \Omega$}\\
0             &\textrm{otherwise} \end{array}\right.
\end{equation}
Imposing periodic boundary conditions to both the density and the Coulomb interaction
kernel in the extended grid, the potential can be calculated according to the
convolution theorem,
\begin{equation}
V_H(\mb{r})=\sum_{\mb{k}} \overline{n}_{2L}(\mb{k}) \overline{v}_c(\mb{k})
e^{i \mb{k} \cdot \mb{r}}
\end{equation}
where $\overline{n}_{2L}(\mb{k})$ and $\overline{v}_c(\mb{k})$ are respectively finite
Fourier integrals of the density and the Coulomb interaction kernel in the extended grid,
\begin{eqnarray}
 \overline{n}_{2L}(\mb{k})& = &\frac{1}{\Omega_{2L}}\int_{\Omega_{2L}}n_{2L}(\mb{r})e^{-i \mb{k} \cdot \mb{r}}d\mb{r},\\
 \overline{v}_c(\mb{k})& = &\int_{\Omega_{2L}}v_c(\mb{r})e^{-i \mb{k} \cdot \mb{r}}d\mb{r}
\end{eqnarray}
where $\Omega_{2L}$ is used to denote both the extend grid and its volume (or area in the 2D case).

While $\overline{n}_{2L}(\mb{k})$ can be easily obtained from the discrete
Fourier transform of its real space values by FFT, the calculation of
$\overline{v}_c(\mb{k})$ is much more involved because of the singularity of
the Coulomb interaction kernel in real space. The key to Martyna-Tuckerman's
approach is to decompose the Coulomb interaction kernel into long- and
short-range parts,
\begin{equation}
v_c(\mb{r})=\frac{\mathrm{erf}(\alpha r)}{r}+\frac{\mathrm{erfc}(\alpha
r)}{r}\equiv v_c^{(long)}(\mb{r})+v_c^{(short)}(\mb{r})
\end{equation}
where $\mathrm{erf}(x)$ and $\mathrm{erfc}(x)$ are the error function and its complement,
respectively, and $\alpha$ is the parameter that controls the effective cut-off range.
The finite Fourier integral of the short range part can be well approximated by its infinite
Fourier transform,
\begin{eqnarray}
\overline{v}_c^{(short)}(\mb{k}) & \equiv & \int_{\Omega_{2L}}
v_c^{(short)}(\mb{r})e^{-i\mb{k}\cdot\mb{r}} d\mb{r} \nonumber \\
& \approx & \int_{\textrm{\small whole space}} v_c^{(short)}(\mb{r})
e^{-i\mb{k}\cdot\mb{r}}d\mb{r} \equiv \tilde{v}_c^{(short)}(\mb{k}). \nonumber \\
\end{eqnarray}
which is analytically known in both 2D and 3D cases. The finite Fourier integral of the
long-range interaction can be directly obtained from the discrete Fourier transform of its
real space values. In the practical implementation, $ \overline{v}_c $ needs to be calculated
only once at the beginning. The calculation of $V_H(\mb{r}) $ involves only two FFTs
(one forward and one backward), which makes this approach much more efficient than
methods based on a Poisson solver.

\subsection{A simplified multi-grid technique}
The multi-grid method is an efficient technique to accelerate the convergence
in various real space relaxation approaches.\cite{NumRecip} The basic idea is
that low-frequency errors are easier to eliminate in a coarse grid than in a
fine grid. Lee et al. \cite{Lee00} proposed a simple one-way multi-grid
technique in finite-difference real space KS calculations. Wave functions being
represented in real-space in our algorithm, a similar simplified multi-grid
(SMG) method can be implemented in a straightforward way: the KS energy
functional is first minimized in a coarse grid; the converged wave functions,
after interpolation and re-orthogonalization, are taken as the initial guess
for the minimization on the fine grid. With these well-preconditioned initial
wave functions, the convergence in the fine grid can be easily attained.

\section{Numerical Tests}

In most experimental QD systems, the excitation in the vertical
direction can be neglected, and a 2D model is a good
approximation. In this paper we will report test results only
in a 2D system; the extension to 3D systems is straightforward,
and the high efficiency of our algorithm makes it possible to
explore large dots even in the 3D case. We test the performance of
the new algorithm in a coupled quartic oscillator potential
system,
\begin{equation}
\label{eq:cqo}
V_\mathrm{ext}(\mb{r})=a\left( \frac{x^{4}}{b}+by^{4}-2\lambda x^{2}y^{2}+
\gamma (x^{2}y-xy^{2})r\right),
\end{equation}
with \( a=10^{-4}\), \(b=\pi/4\), \(\lambda =0.6\) and \(\gamma =0.1\).
Considering that the convergence behavior in KS calculations is generally related
to the effective interaction strength, which is characterized
by \( r_{s}\equiv (\pi \overline{n})^{-1/2}/a_{B} \) in the 2D case
where \( \overline{n} \) is the average density and \( a_{B} \)
is the Bohr radius, we have chosen the parameters in $V_\mathrm{ext}$ so that
the estimated \( r_{s} \) is about 1.5, which is close to experimental values.
The calculations are done in a grid of size $L_x=L_y=50$ and
the number of grid points is $N_x=N_y=64$. All the following numerical results
come from calculations with electron number \( N=100 \) and spin $S=0$.
For the exchange-correlation energy, $E_\mathrm{xc}$,
we use Tanatar and Ceperley's parameterized form of the LSDA functional. \cite{Tanatar89}
The convergence criterion is set as $\epsilon=10^{-6}$, which corresponds to about
$10^{-5}$ meV in GaAs-AlGaAS QD systems.

\begin{figure}
\includegraphics[width=2.8in]{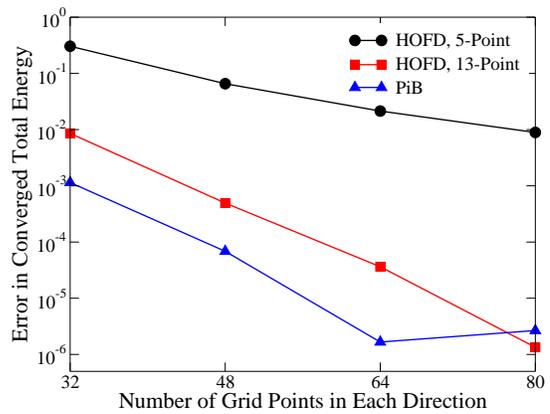}
\caption{\label{fig:grid}
Convergence of the total energy with respect to the number of grid points using 5-Point FD,
13-Point FD and PiB representation.}
\end{figure}

Considering that the FD method has been widely used in the numerical modelling
of QD systems,\cite{Kumar90,Macucci93,Lee98,Pi01} we first make a comparison
between FD and PiB. In the FD representation, the second order derivative in
the kinetic energy operator is locally discretized in real space
\begin{equation}
\label{eq:FD}
\frac{\partial^2}{\partial x^2}f(x)=\frac{1}{h^2} \sum_{j=-m}^{m}C_j f_j +\mathrm{O}(h^{2m})
\end{equation}
where $h$ is the discretization step and the coefficients \(C_j\)
can be obtained systematically for any $m$. Fig. \ref{fig:grid}
illustrates the convergence of the total energy with respect to
the number of grid points using 5-point ($m=2$), 13-point ($m=6$)
FD and PiB representations. The lower-order finite difference
scheme, $m=2$, is poor in terms of accuracy, and converges slowly
as the grid size increases. The high-order FD scheme, $m=6$, does
improve the accuracy by two orders of magnitude, but is still less
accurate than PiB for $N_x=32$, $48$ and $64$.


\begin{figure}
\includegraphics[width=2.8in]{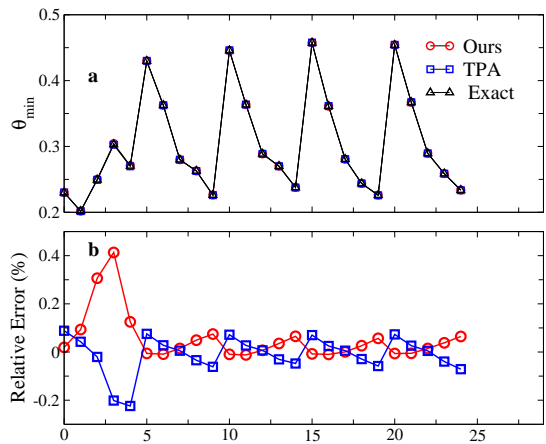}
\caption{\label{fig:linmin}
Comparison of $\theta_\mathrm{min}$ calculated by three different schemes.}
\end{figure}

We check the accuracy and efficiency of our line-minimization scheme by
comparing with TPA's approach as well as the numerically exact Brent's line
search algorithm.\cite{NumRecip} In this case, we use $N_\mathrm{band}=5$ and
$N_\mathrm{update}=1$ as recommended in TPA's original algorithm.\cite{Payne92}
Fig. \ref{fig:linmin} plots $\theta_\mathrm{min} $ in the first 25 band
iterations calculated from three schemes respectively. The values of
$\theta_\mathrm{min} $ calculated from both TPA's and our method agree very
well with exact values and the relative errors are always  smaller than $1\% $.
But in terms of computational effort, our new scheme is much more efficient as
argued in the previous section.


\begin{figure}
\includegraphics[width=3.0in,clip]{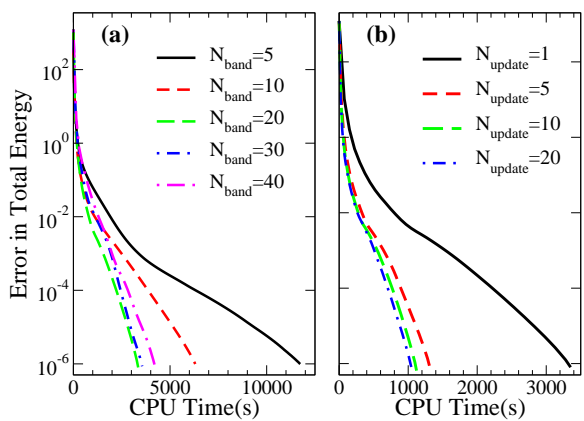}
\caption{\label{fig:Nband}
Errors in the total energy as a function of CPU time during the DMCG calculation
for different $N_\mathrm{band}$ with $N_\mathrm{update}=1$ (a) and
for different  $N_\mathrm{update}$ with $N_\mathrm{band}=20$ (b) with
respect to the exact total energy that is calculated using a finer grid ($N_x=80$)
and tighter convergence criterion ($\epsilon=10^{-7}$) ,}
\end{figure}

To find the optimal \( N_\mathrm{band} \) and \( N_\mathrm{update} \),
we do the calculations with different values of \( N_\mathrm{band} \) and \( N_\mathrm{update}\),
and the results are shown in Fig. \ref{fig:Nband}. With fixed \(N_\mathrm{update}=1\), it is seen
that a relatively larger \( N_\mathrm{band} \) is more efficient than small \(N_\mathrm{band}\).
Fixing \(N_\mathrm{band}=20\), \(N_\mathrm{update}=20\) gives the best performance. The combination of
large \(N_\mathrm{band}\) and \(N_\mathrm{update}\) reduces the computational effort by almost one order
of magnitude. Though the actual values of optimal \( N_\mathrm{band} \) and \( N_\mathrm{update} \)
may vary for different systems, the basic idea demonstrated in this test calculation is believed to
be of general significance.

\begin{figure}
\includegraphics[width=3.0in,clip]{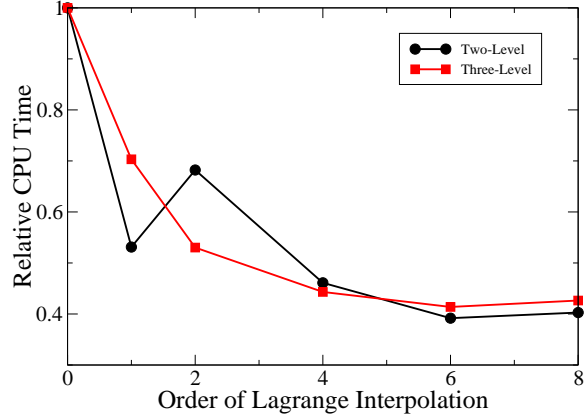}
\caption{\label{fig:smg}
Relative computational efforts for one KS calculation as a function of the order of 
interpolation in both two- and three-level SMG calculations. The CPU time in the 
single-level calculation is taken as the unit.}
\end{figure}

We have implemented both the two-level ($h$ and $2h$) and three-level
($h$,$\frac{4}{3}h$ and $2h$) SMG schemes. Instead of using a sophisticated
interpolation as in Ref. \onlinecite{Lee00}, we use a simpler
Lagrange polynomial interpolation method. Comparison with the single-level calculation
shows a quite obvious improvement in computational efficiency. To check the effect
of the interpolation accuracy, in Fig. \ref{fig:smg} we plot the relative computational 
effort in one KS calculation as a function of the order of the Lagrange interpolation 
formula in both two- and three- level SMG calculations. We see that a high-order interpolation 
scheme is useful to improve the performance.

\section{Summary}
  In this paper, we presented an efficient algorithm for the KS-SDFT simulation of
large quantum dot systems. The main elements of the algorithm
are: (1) Wave functions are represented in real space, and the
kinetic energy operator is applied to wave functions by fast sine
transform. (2) The Hartree potential is calculated by
Martyna-Tuckerman's Fourier convolution method. (3) For the
solution of the KS equation, we introduced several important
modifications to Teter et al.'s band-by-band conjugate-gradient
method. A more efficient approximate line-minimization scheme was
developed; it was found that large band iteration number and a
delayed update of the KS Hamiltonian inside the band iterations
increase the efficiency by one order of magnitude. (4) A
simplified multi-grid technique was introduced to accelerate the
convergence. The new algorithm has been used to study spin and
conductance peak-spacing distributions in a 2D chaotic QD system
with electron number $N$ up to 200, from which new physical
phenomena were revealed.\cite{Jiang02}

\begin{acknowledgments}
This work was supported in part by NSF Grant No. DMR-0103003.
\end{acknowledgments}


\end{document}